\documentclass[aps,pra,superscriptaddress,amsmath,amssymb,preprintnumbers,twocolumn]{revtex4}
\usepackage{amssymb}
\usepackage{booktabs}
\usepackage{graphicx}
\usepackage{bm}
\usepackage{color}
\usepackage{cancel}
\usepackage{slashed}
\usepackage{subfigure}

\newcommand{\Ignore}[1]{}

\newcommand{\Ket}[1]{\left\vert #1\right\rangle}
\newcommand{\Bra}[1]{\left\langle #1\right\vert}

\newcommand{\kboltz}{k_\mathrm{B}}
\newcommand{\psiup}{\Ket{1}}
\newcommand{\psizero}{\Ket{0}}
\newcommand{\psidown}{\Ket{-1}}

\newcommand{\ii}{\mathrm{i}}
\newcommand{\ee}{\mathrm{e}}

\begin{document}

\title{Master equation approach to the three-state open Majorana model}

\author{Benedetto Militello}
\address{Universit\`a degli Studi di Palermo, Dipartimento di Fisica e Chimica - Emilio Segr\`e, Via Archirafi 36, 90123 Palermo, Italia}
\address{I.N.F.N. Sezione di Catania, Via Santa Sofia 64, I-95123 Catania, Italia}

\author{Nikolay V. Vitanov}
\address{Department of Physics, Sofia University, James Bourchier 5 Boulevard, 1164 Sofia, Bulgaria}

\begin{abstract}
The three-state Majorana model in the presence of dissipation is considered. Different models of system-environment interaction are explored, ranging from situation where dissipation is the main effect to regimes where dephasing is mainly produced. It is shown that the detrimental effects of the noise are stronger in the presence of dissipation than in the presence of dephasing.  The role of temperature is also discussed.
\end{abstract}

\maketitle

\section{Introduction}\label{sec:introduction}

Analytical resolution of Schr\"odinger problems with non static Hamiltonians are rare and restricted to specific situations~\cite{ref:Barnes2012,ref:Simeonov2014,ref:Sinitsyn2018,ref:Sriram2005,ref:Owusu2010,ref:Chrusc2015}.
The Landau-Zener-Majorana-Stueckelberg (LZMS) model~\cite{ref:Landau,ref:Zener,ref:Majo,ref:Stuck}, introduced in 1932 independently by four scientists, is a noteworthy and very famous time-dependent model allowing an explicit calculation of the propagator from $t=-\infty$ to $t=+\infty$.  The Majorana's formulation of the problem is more general than the formulations of the other three authors, since it involves a multi-state system instead of a two-state one.
The two-state version of the problem has been extensively studied relaxing several assumptions of the original formulation, hence considering non linear time-dependence of the bare energies~\cite{ref:Vitanov1999b} or a finite duration of the expertiment~\cite{ref:Vitanov1996,ref:Vitanov1999a}. Unusual extensions involving nonlinear equations~\cite{ref:Ishkhanyan2004,ref:MilitelloPRE2018} or non-Hermitian Hamiltonians~\cite{ref:Toro2017} have also been analyzed.
In some studies, the original LZMS model characterized by an avoided crossing has been replaced by the hidden crossing model~\cite{ref:Fishman1990,ref:Bouwm1995}, where neither the bare nor the dressed energies cross, or by the total crossing model~\cite{ref:Militello2015a}, where both the bare and the dressed energies cross. Also variants of the multi-state LZMS have been studied, by considering two degenerate crossing levels~\cite{ref:Vasilev2007} or many levels which cross according to different schemes: the Demkov-Osherov (or equal-slope) model~\cite{ref:Demkov1967}, which consists in a sequence of binary crossings that can be treated with the Independent Crossing Approximation~\cite{ref:Brundobler1993}, or the bowtie model~\cite{ref:Carroll1986a,ref:Carroll1986b,ref:Demkov2000}, where a proper multi-level crossing of the bare energies occur.
Other specific situations involving precise numbers of states with specific coupling schemes have been introduced and solved~\cite{ref:Shytov2004,ref:Sin2015,ref:Li2017,ref:Sin2017}.
Multi-level LZMS transitions are present in many physical scenarios, including very important spin-boson systems described by the time-dependent Rabi Hamiltonian~\cite{ref:Dodo2016} or the Tavis-Cummings model~\cite{ref:Sun2016,ref:Sin2016}.

In order to make the analysis of the LZMS models more realistic, also decoherence and dissipation induced by the interaction with the environment have been considered. Beyond general treatments of quantum noise in adiabatic or quasi adiabatic evolutions~\cite{ref:Lidar,ref:Florio,ref:ScalaOpts2011,ref:MilitelloPScr2011,ref:Wild2016}, specific studies on open two-state systems described by LZMS models have been reported~\cite{ref:Ao1991,ref:Potro2007,ref:Wubs2006,ref:Saito2007,ref:Lacour2007,ref:Nel2009,ref:ScalaPRA2011}. Concerning the multi-state version, there are only few contributions. Ashhab~\cite{ref:Ashhab2016} has analyzed a class of multi-level LZMS problem in the presence of quantum noise, mainly focusing on environment-induced dephasing. Very recently, a three-state LZMS problem in the presence of decays toward external states has been considered~\cite{ref:Militello2019a,ref:Militello2019b}. 

The three-state Majorana model can be thought of as (but is not limited to) the problem of a spin-1 particle in a time-dependent magnetic field whose $z$-component linearly changes with time, while the $x$-component is static. The specific analysis of a three-state Majorana model can be considered a trivial extension of the two-state counterpart, especially because it is well known that the three-state Majorana model can be reduced to an effective two-state LZMS model. In spite of this, it is worth observing that when the system is subjected to noise it is not possible to obtain an effective two-state model, as discussed in the following.
Over the decades, some papers appeared dealing with the Majorana model in the presence of noise. Ellinas has considered an Hamiltonian with a phenomenological non-Hermitian term proportional to the longitudinal component of the angular momentum~\cite{ref:Ellinas1992}, while other Authors such as Kenmoe {\it et al.}~\cite{ref:Kenmoe2013} have considered the effects of a classical colored noise due to fluctuations of the parameters of the classical fields. Very recently, Band and Avishai have considered a phenomenological master equation to describe the effects of a white Gaussian isotropic noise~\cite{ref:Band2019}. An analysis of the quantum noise based on a microscopic model has been developed almost two decades ago by Saito and Kayanuma~\cite{ref:Saito2002}, but essentially limited, for the spin-$1$ case, to interaction operators which are diagonal in the basis of the diabatic states, and to colored or strong white noise. Pokrovsky and Sinitsyn have generalized this treatment to general white noise and transverse system-environment couplings~\cite{ref:Pokrovsky2003}. 

In this paper we report an new analysis of the effects of quantum noise on the Majorana model. Differently from Refs.~\cite{ref:Saito2002,ref:Pokrovsky2003}, our mathematical treatment is based on the Davies and Spohn theory~\cite{ref:DaviesSpohn}, which allows for deriving the correct master equation for systems governed by time-dependent Hamiltonians in the presence of Markovian noise.
Moreover, we take into account both contributions inducing dissipation and interaction terms mainly associated with dephasing processes, all related to the same bath, exploring the transition from one regime to the other and showing that dissipation is more detrimental than dephasing, though not dramatically worse. 
The paper is structured as follows. In the next section we introduce the ideal spin-1 Majorana model and its main features. In section~\ref{sec:dissipative} we introduce the interaction with an environment and derive the relevant master equation. In section~\ref{sec:efficiency} we evaluate the efficiency for different schemes of system-environment interaction, first focusing on the zero-temperature case and then including the thermal effects. Finally, in section~\ref{sec:conclusions} we give some conclusive remarks.

\section{Three-State Majorana Model}\label{sec:ideal}

We consider a three-state system with a cascade coupling scheme and two linearly time-dependent bare energies ($\hbar=1$):
\begin{eqnarray}\label{eq:Hamiltonian_ideal_matrix}
\hat{H}(t) = \left(
\begin{array}{ccc}
-\kappa t & \Omega/\sqrt{2} & 0 \\
\Omega/\sqrt{2} & 0 & \Omega/\sqrt{2} \\
0 & \Omega/\sqrt{2} & \kappa t
\end{array}
\right)\,,
\end{eqnarray}
which is a spin-$1$ Majorana model:
\begin{subequations}
\begin{eqnarray}\label{eq:Hamiltonian_ideal}
\nonumber
\hat{H}(t) = \kappa t \hat{J}_z + \Omega \hat{J}_x = \omega(t) \hat{J}_\theta\,,
\end{eqnarray}
with
\begin{eqnarray}
\hat{J}_\theta &=& \cos\theta \hat{J}_z + \sin\theta \hat{J}_x \,, \\
\theta &=& \pi/2 -\arctan[\kappa t  / \Omega] \,, \\
\omega(t) &=& \sqrt{(\kappa t)^2+\Omega^2}\,.
\end{eqnarray}
\end{subequations}

This model can be related to many situations, beyond a spin-1 particle in a non-static magnetic field. For example, it can be easily implemented with artificial atoms~\cite{ref:NoriPRL2005,ref:NoriSR2014}.

This Hamiltonian can be diagonalized through the action of the unitary operator $\hat{U}_y(\theta) = \ee^{\ii \theta \hat{J}_y}$. Since we have $\hat{U}_y(\theta) \hat{H}(t) \hat{U}_y(-\theta) = \omega(t) \hat{J}_z$, the eigenvalues of $H$ are $-\omega(t)$, $0$ and $\omega(t)$, while its eigenstates can be obtained as $\Ket{1,m}_\theta = \hat{U}_y(-\theta) \Ket{1,m}_z$, then obtaining, respectively (and simplifying the notation as $\Ket{m}\equiv\Ket{1,m}$),
\begin{subequations}
\begin{eqnarray}
\nonumber
\psidown_\theta &=& \frac{1+\cos\theta}{2}\psidown_z - \frac{\sin\theta}{\sqrt{2}}\psizero_z  + \frac{1-\cos\theta}{2}\psiup_z  \,, \\
&& \\
\nonumber
\psizero_\theta &=& \frac{\sin\theta}{\sqrt{2}}\psidown_z + \cos\theta\psizero_z - \frac{\sin\theta}{\sqrt{2}}\psiup_z  \,, \\
&& \\
\nonumber
\psiup_\theta &=&  \frac{1-\cos\theta}{2}\psidown_z + \frac{\sin\theta}{\sqrt{2}}\psizero_z + \frac{1+\cos\theta}{2}\psiup_z   \,.\\
\end{eqnarray}
\end{subequations}

It is easy to see that the eigenstate $\psiup_\theta$ approaches $\psidown_z$ for $t\rightarrow -\infty$ ($\theta\rightarrow \pi$) and $\psiup_z$ for $t\rightarrow \infty$  ($\theta\rightarrow 0$). Then it can be used for realizing a population transfer from the first to the third state. Analogously, $\psidown_\theta$ realizes the transfer from $\psiup_z$ to  $\psidown_z$.
In the ideal case and in the adiabatic limit, the population transfer is practically perfect.
The third adiabatic state, $\Ket{0}_\theta$, is a linear combination of the states $\Ket{0}_z$, with coefficient $\cos\theta$, and $(\Ket{-1}_z-\Ket{1}_z)/\sqrt{2}$, with coefficient $\sin\theta$, so that when $t$ goes from $-\infty$ to $0$, the angle $\theta$ varies from $\pi$ and $\pi/2$, making the adiabatic state varies from $\Ket{0}_z$ to  $(\Ket{-1}_z-\Ket{1}_z)/\sqrt{2}$. This circumstance paves the way to the possibility of describing the three-state Majorana model as an effective two-state LZMS model, as in Ref.~\cite{ref:Randall2018}.

\section{The Open Majorana Model}\label{sec:dissipative}

The standard theory of open quantum systems for time-independent Hamiltonians, in the presence of Markovian noise, describes transitions between the instantaneous eigenstates of the system Hamiltonian due to the system-environment interaction~\cite{ref:Petru,ref:Gardiner}.  Assuming a system-environment interaction Hamiltonian
\begin{eqnarray}
H_I = \lambda \, \hat{X} \otimes \hat{B}\,,
\end{eqnarray}
these processes are described by the jump operators $\hat{X}(\nu) = \sum_{\epsilon'-\epsilon=\nu} \hat{\Pi}_\epsilon \, \hat{X} \, \hat{\Pi}_{\epsilon'}$, with $\hat{\Pi}_\epsilon $ the instantaneous eigenprojectors of the Hamiltonian, i.e., $\hat{\Pi}_\epsilon \hat{H} = \hat{H}\hat{\Pi}_\epsilon = \epsilon\hat{\Pi}_\epsilon$. Irrespectively of the structure of $\hat{X}$, the jump operators connect the eigenspaces of the Hamiltonian.

When the Hamiltonian is time-dependent and the noise is memoryless, one can analyze the dissipative dynamics following to the Davies and Spohn theory~\cite{ref:DaviesSpohn}. According to such approach based on the assumption that the bath correlation time is virtually zero, hence much smaller than the time scale on which the Hamiltonian changes, one can analyze the system by considering short time windows where the Hamiltonian is essentially time independent and the environment is responsible for a Markovian noise. The relevant master equation describes a dissipative dynamics that connects the instantaneous eigenspaces of the Hamiltonian. Indeed, the terms related to such transitions are the only ones surviving the secular approximation performed in the derivation of the master equation. Applications of this theory can be found for example in Refs.~\cite{ref:Florio,ref:ScalaOpts2011,ref:MilitelloPScr2011,ref:MilitelloPRA2010}. Assuming a weak system-bath coupling and a flat spectrum (to guarantee short bath correlation times) one obtains the following Markovian master equation:
\begin{subequations}
\begin{eqnarray}
\nonumber
\dot\rho &=& -\ii [H(t), \rho] + \sum_{\nu} \gamma(\nu) [ \,\, \hat{X}(\nu) \rho \hat{X}^\dag(\nu) \\
\nonumber
&-& \frac{1}{2}\{  \hat{X}^\dag(\nu)\hat{X}(\nu), \rho \}  ]  \,, \qquad \nu=\pm\omega(t), \pm 2\omega(t)\,, \\
\end{eqnarray}
where $\lambda^2 [N(\nu, T) + 1]$ for positive frequencies and $\lambda^2 N(|\nu|, T)$ for negative values of $\nu$, with $N(\nu,T)=[\ee^{\nu/(\kboltz T)}-1]^{-1}$ the mean number of bath bosons at frequency $\nu$, being $T$ the bath temperature and $\kboltz$ the Boltzmann constant. Moreover, we have
\begin{eqnarray}
\hat{X}(2\omega(t))&=&  \hat{P}_{-1}(\theta) \, \hat{X} \, \hat{P}_1(\theta) \,,  \\
\hat{X}(\omega(t))&=& \sum_{m=-1,0} \hat{P}_m(\theta) \, \hat{X} \,  \hat{P}_{m+1}(\theta) \,, \\
\hat{X}(-\omega(t))&=& \sum_{m=0,1}  \hat{P}_m(\theta) \, \hat{X} \,  \hat{P}_{m-1}(\theta) \,, \\
\hat{X}(-2\omega(t))&=& \hat{P}_{1}(\theta) \, \hat{X} \,  \hat{P}_{-1}(\theta) \,.
\end{eqnarray}
\end{subequations}
with $\hat{P}_m(\theta) =  \Ket{m}_\theta \Bra{m}$.

It is worth observing that in the presence of dissipation, incoherent transitions involving the three adiabatic states $\psidown_\theta$, $\psizero_\theta$ and $\psiup_\theta$ are induced and it is not possible anymore to consider an effective two-state model, as can be done in the ideal case. In fact, specifically for the example considered in the previous section, dissipation will produce transitions from $\psizero_\theta$ to the other states, and a dynamics involving only $\Ket{0}_z$ and  $(\Ket{-1}_z-\Ket{1}_z)/\sqrt{2}$ will not be possible anymore.

We conclude this section by commenting on the assumption of Markovian quantum noise, based on the hypotheses of weak coupling and flat spectrum. This second condition has to be meant as flatness of the spectrum in the frequency range of the system transition frequencies. Moreover, in standard situations non-Markovian behaviors typically emerge in the short-time dynamics, unless the environment is suitably \lq structured\rq\, to maintain memory in a wider time interval. Since the adiabatic following that realizes the population transfer concerns a long time interval, the non-Markovian behaviors are quickly lost in favor of a Markovian behavior; because of the weak coupling, no significant effects are produced during the non-Markovian evolution. Therefore, in our case, it is quite reasonable to assume Markovianity, though exploration of non-Markovian behaviors could be of interest in order to include a wider range of physical situations. For example, protocols exploiting shortcuts to adiabaticity, then involving shorter time intervals, may require a different treatment aimed at describing non-Markovian effects. But these scenarios are beyond the scope of the present work.

\section{Efficiency of the population transfer}\label{sec:efficiency}

Let us assume
\begin{eqnarray}
\hat{X} = \cos\phi \hat{J}_z + \sin\phi \cos\varphi \hat{J}_x + \sin\phi \sin\varphi \hat{J}_y\,,
\end{eqnarray}
and evaluate the efficiency for different values of the system-environment coupling strength $\lambda$ and of the two angles $\phi$ and $\varphi$ defining the operator $\hat{X}$. We also consider the effects of temperature. The efficiency is always evaluated as the population of the target state in the final state of the system, whose dynamics is obtained through a numerically exact resolution of the relevant Schr\"odinger equation.

We observe that whatever the angles $\phi$ and $\varphi$ are, we always have a dissipative dynamics, not a pure dephasing. This is due to the fact that the commutator between $\hat{X}$ and $H$ is generally non-zero, and can vanish only at specific instants of time. Nevertheless, when $|\kappa t|\gg \Omega$ the eigenstates of $H$ are very close to the those of $\hat{J}_z$, which means that under such condition, if $\hat{X}\approx \hat{J}_z$, the system mainly undergoes dephasing, and only minor dissipation effects. Of course, in the proximity of $t=0$ it happens that $H$ is essentially proportional to $\hat{J}_x$ and an operator $\hat{X}\approx \hat{J}_z$ mainly induces dissipation.
On the contrary, if $\hat{X}\approx\hat{J}_x$, the system mainly undergoes dissipation, while in the proximity of $t=0$ it is prominently subjected to dephasing.

\begin{figure}[h!]
\begin{tabular}{cl}
\subfigure[]{\includegraphics[width=0.38\textwidth, angle=0]{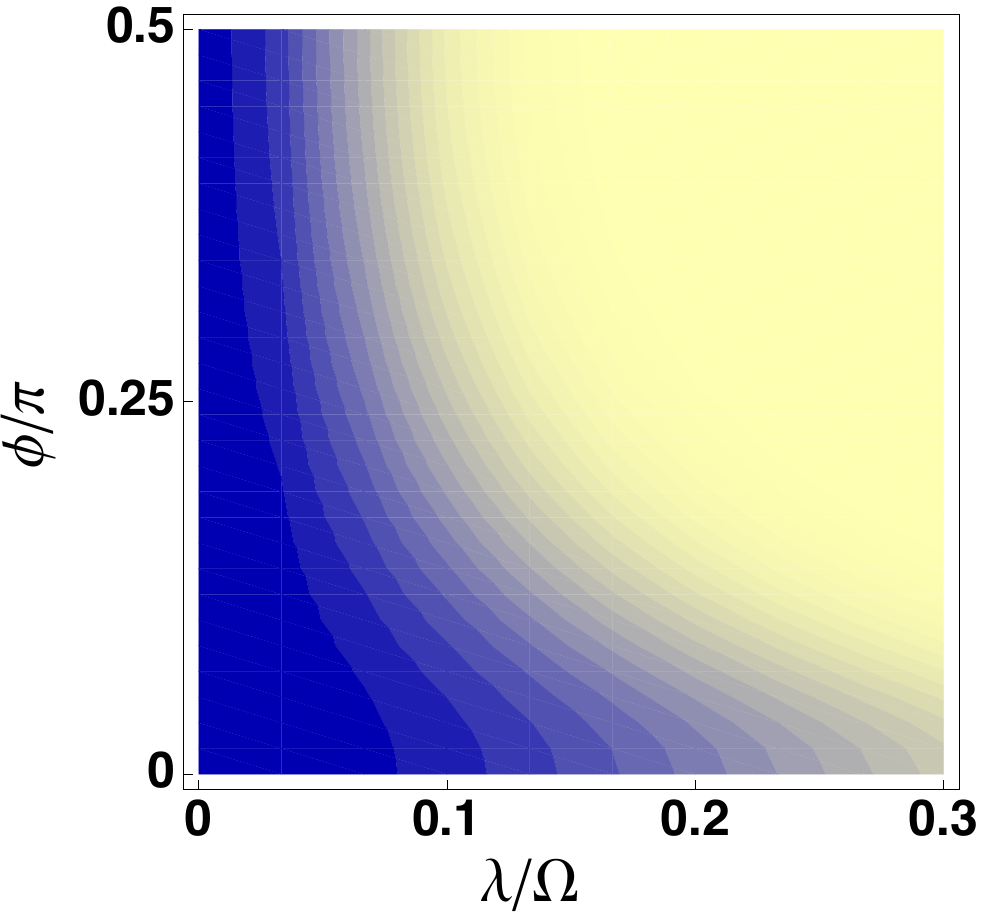}} &
\includegraphics[width=0.07\textwidth, angle=0]{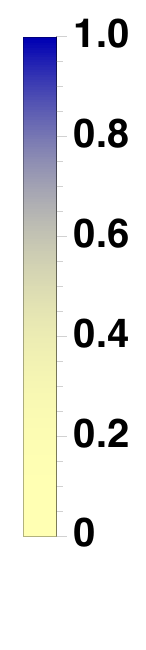} \\
\subfigure[]{\includegraphics[width=0.38\textwidth, angle=0]{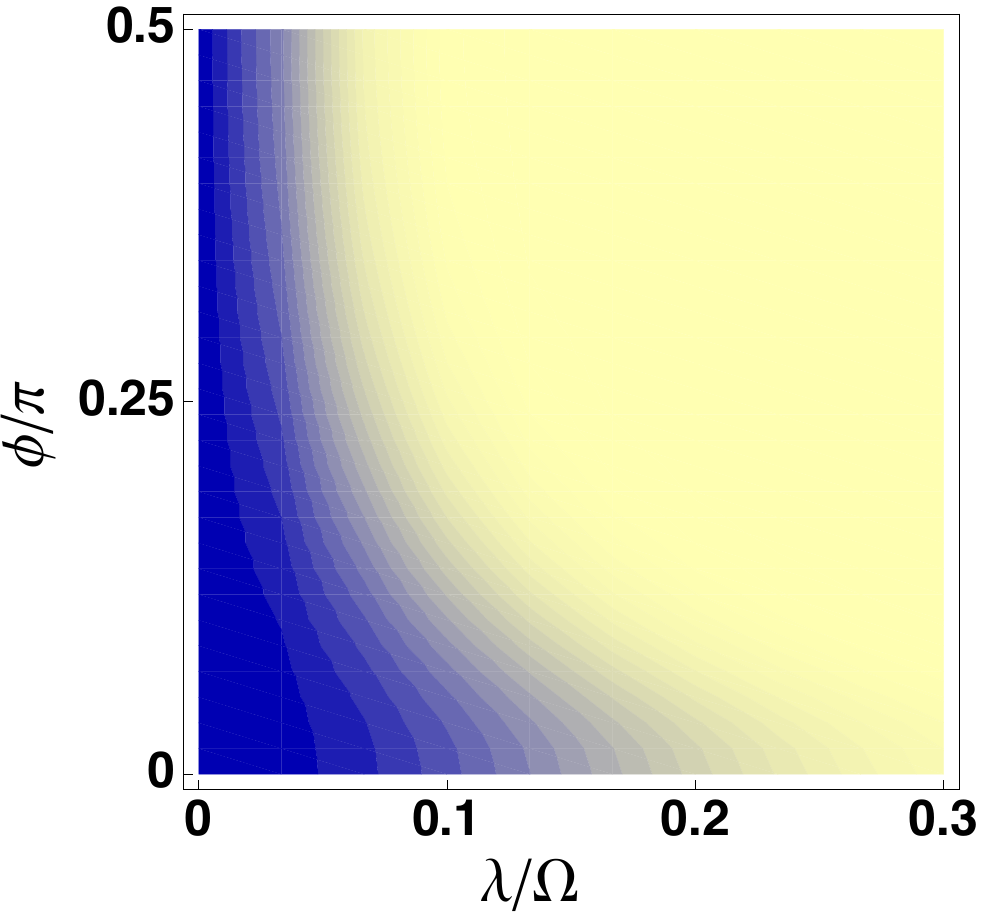}} &
\includegraphics[width=0.07\textwidth, angle=0]{fL1.pdf} \\
\end{tabular}
\caption{(Color online) Final population of state $\psiup_z$ when the system starts in $\psidown_z$ as a function of $\lambda/\Omega$ and $\phi/\pi$, at essentially zero temperature ($\kboltz T/\Omega = 0.001$).  In (a) $\kappa/\Omega^2=0.25$ while in (b)  $\kappa/\Omega^2=0.1$. The other relevant parameters are: $\varphi=0$ and $\kappa t_0 / \Omega = 25$, both in (a) and (b).} \label{fig:diss-1}
\end{figure}

\begin{figure}[h!]
\begin{tabular}{cl}
\subfigure[]{\includegraphics[width=0.38\textwidth, angle=0]{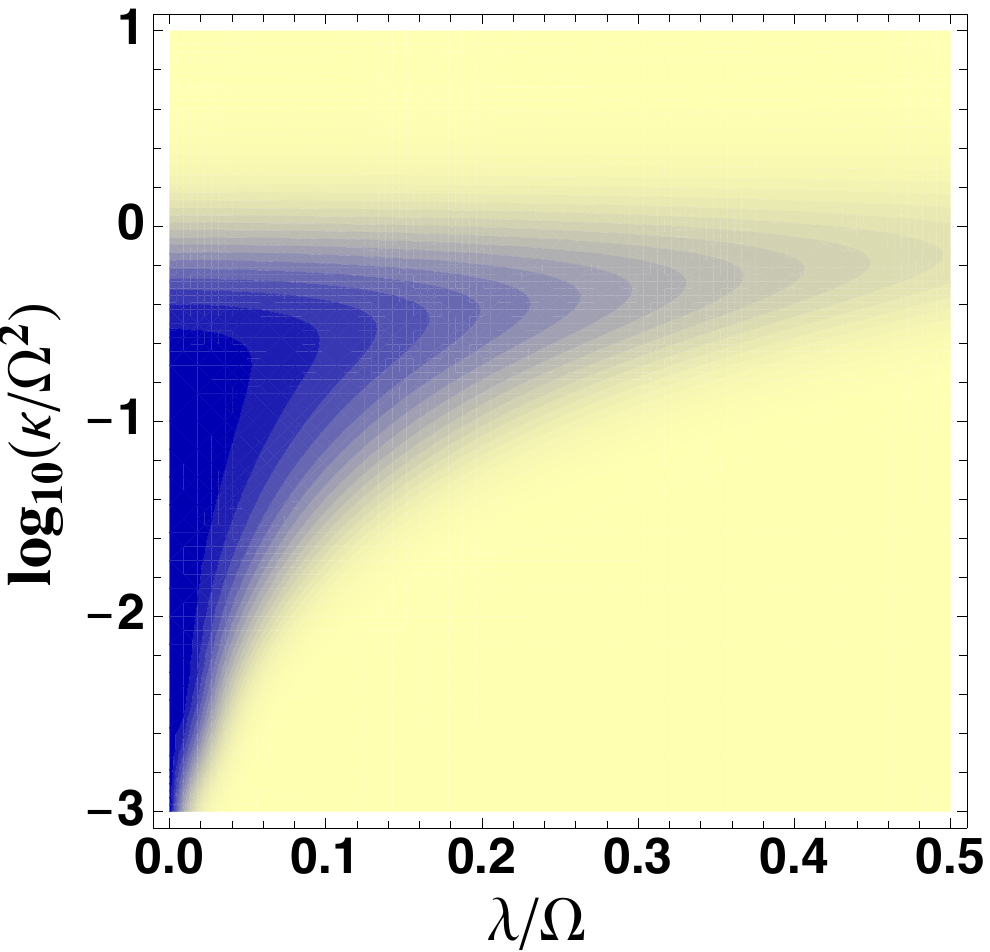}} &
\includegraphics[width=0.07\textwidth, angle=0]{fL1.pdf} \\
\subfigure[]{\includegraphics[width=0.38\textwidth, angle=0]{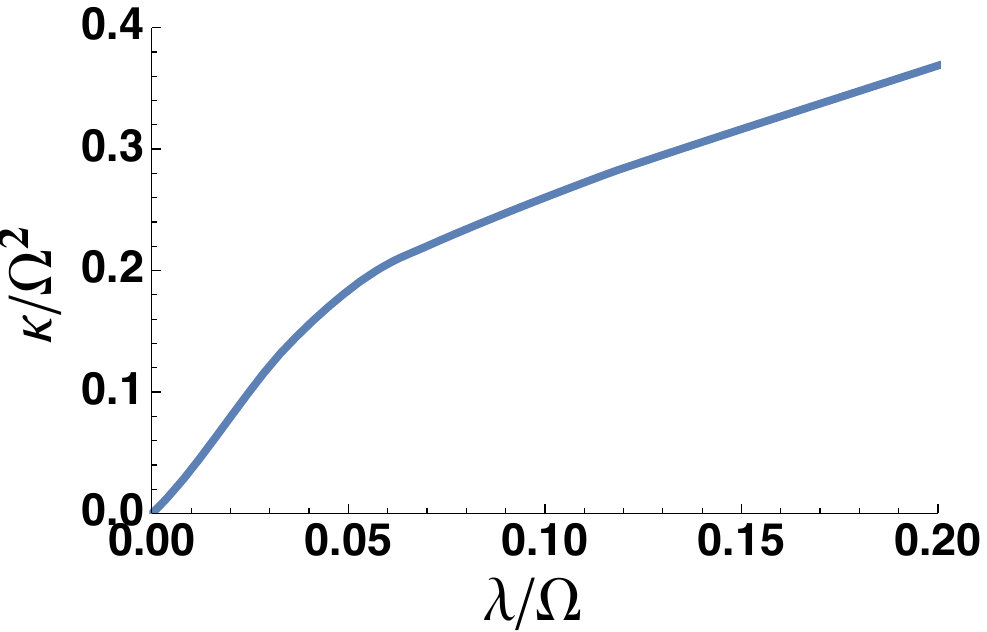}} &  \\
\end{tabular}
\caption{(Color online) In (a) it is plotted the final population of state $\psiup_z$ when the system starts in $\psidown_z$ as a function of $\lambda/\Omega$ and $\kappa/\Omega^2$ (in logarithmic scale). The relevant parameters are: $\kboltz T/\Omega = 0.001$ (essentially zero temperature), $\kappa t_0 / \Omega = 25$, $\varphi=0$ and $\phi=0$ (which imply $\hat{X}=\hat{J}_z$).
In (b) we show the optimal value of $\kappa$ (in units of $\Omega^2$) as a function of $\lambda/\sqrt{\Omega}$ in the range $[0,0.2]$.}\label{fig:kappa-z}
\end{figure}

\begin{figure}[h!]
\begin{tabular}{cl}
\includegraphics[width=0.38\textwidth, angle=0]{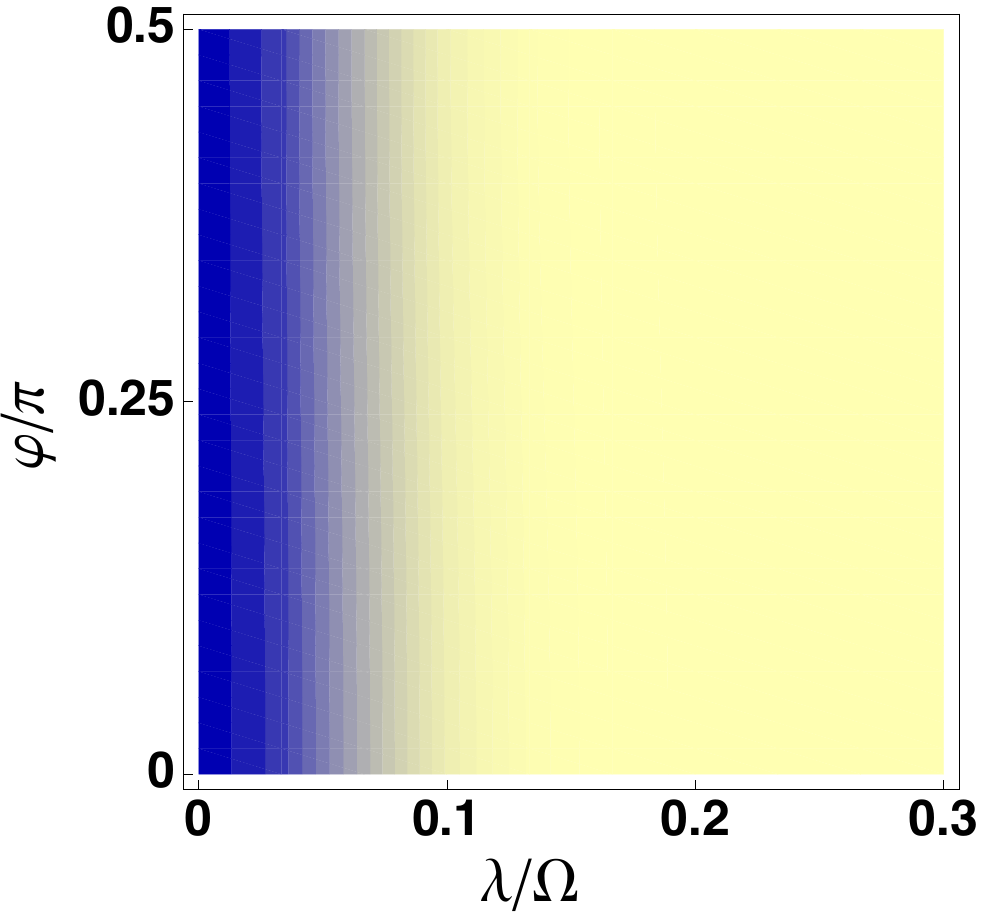} &
\includegraphics[width=0.07\textwidth, angle=0]{fL1.pdf}
\end{tabular}
\caption{(Color online) Final population of state $\psiup_z$ when the system starts in $\psidown_z$ as a function of $\lambda/\Omega$ and $\varphi/\pi$, at essentially zero temperature ($\kboltz T/\Omega = 0.001$). The relevant parameters are: $\phi=\pi/2$, $\kappa / \Omega^2 = 0.25$ and $\kappa t_0 / \Omega = 25$.} \label{fig:efficiency1}\label{fig:diss-2}
\end{figure}


{\it Zero-Temperature Bath --- } In Fig.~\ref{fig:diss-1} the efficiency of the population transfer from $\psidown$ to $\psiup$ as a function of $\phi$ and $\lambda$, assuming $\varphi=0$, therefore spanning linear combinations of $\hat{J}_z$ and $\hat{J}_x$ with different weight for such two operators.
From Fig.~\ref{fig:diss-1}a we can see that the system turns out to be a bit more robust against environmental effects in the limit $\hat{X}\approx \hat{J}_z$ ($\phi\approx 0$) than in the opposite case $\hat{X}\approx \hat{J}_x$ ($\phi\approx \pi/2$).

In Fig.~\ref{fig:diss-1}b a smaller value of $\kappa$ is considered, with a consequent diminishing of the high-efficiency region, due to the exposure of the system to the environmental effects for a longer time. Generally speaking, high values of $\kappa$ must be considered carefully, because they could imply failure of adiabaticity, but very small values should be avoided as well, to reduce the influence of noise.

To better express this concept, in Fig.~\ref{fig:kappa-z}a we have shown the efficiency as a function of $\lambda$ and $\kappa$ in the case $\hat{X}\approx \hat{J}_z$. In a wide range of $\lambda$ values there is a range of values of $\kappa$ where the efficiency is close to unity. While the upper limit is more or less the same (say, $\kappa\sim\Omega$), the lower bound for a good efficiency turns out to increase with $\lambda$. Over a certain value of $\lambda$ the efficiency is always very low. In Fig.~\ref{fig:kappa-z}b we plot the (optimal) value of $\kappa$ that gives the highest efficiency for a given value of $\lambda$. For $\lambda\rightarrow 0$ the optimal value of $\kappa$ approaches zero, but never reaches such a value. Even in the ideal case, $\kappa$ can be as small as one wishes but not zero, to prevent the Hamiltonian to become stationary and the adiabatic following to be lost.

It is interesting to consider also the implications of having a system-environment interaction related to $\hat{J}_y$, which never commutes with $H(t)$, neither exactly nor approximately (we could address this case as the purely dissipative one). In this situation, the system undergoes dissipation during the whole experiment duration $2 t_0$. We have therefore studied the case where $\hat{X}=\cos\phi \hat{J}_x + \sin\phi \hat{J}_y$. In Fig~\ref{fig:diss-2} it is reported the relevant efficiency. Independence from the parameter $\phi$ is evident.

\begin{figure}[h!]
\includegraphics[width=0.45\textwidth, angle=0]{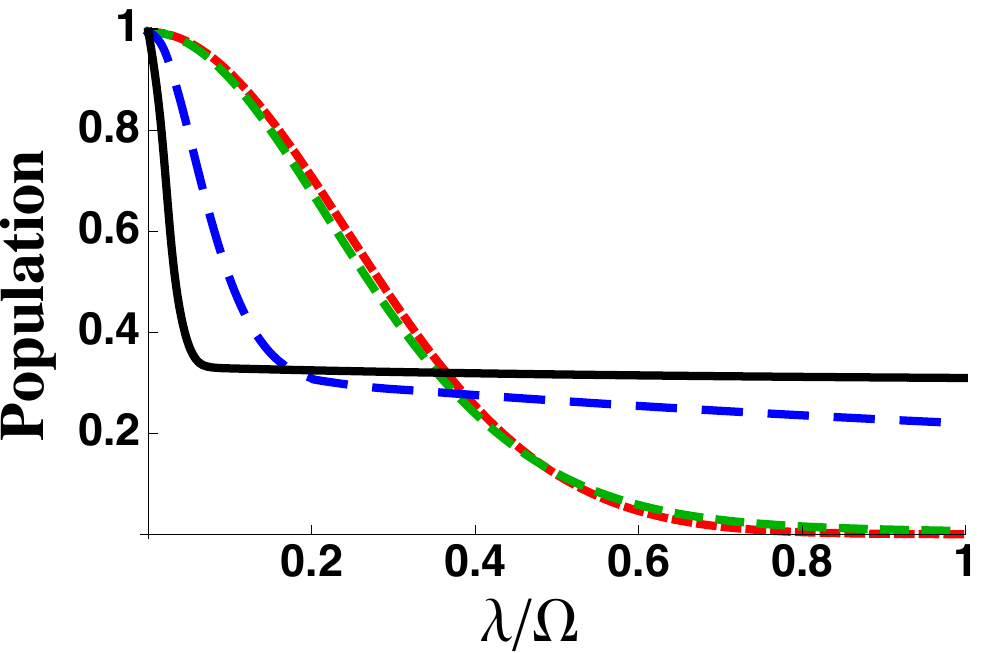}
\caption{(Color online) Final population of state $\psiup_z$ when the system starts in $\psidown_z$ as a function of $\lambda/\Omega$, for different temperatures: $\kboltz T/\Omega = 0.001$ (red dotted line), $\kboltz T/\Omega = 1$ (green dashed line, very close to the red dotted one), $\kboltz T/\Omega = 3$ (blue long-dashed line) and $\kboltz T/\Omega = 10$ (black solid line). The relevant parameters are: $\kappa / \Omega^2 = 0.25$, $\kappa t_0 / \Omega = 25$, $\varphi=0$ and $\phi=0$ (which imply $\hat{X}=\hat{J}_z$).}\label{fig:temp-z}
\end{figure}
\begin{figure}[h!]
\subfigure[]{\includegraphics[width=0.45\textwidth, angle=0]{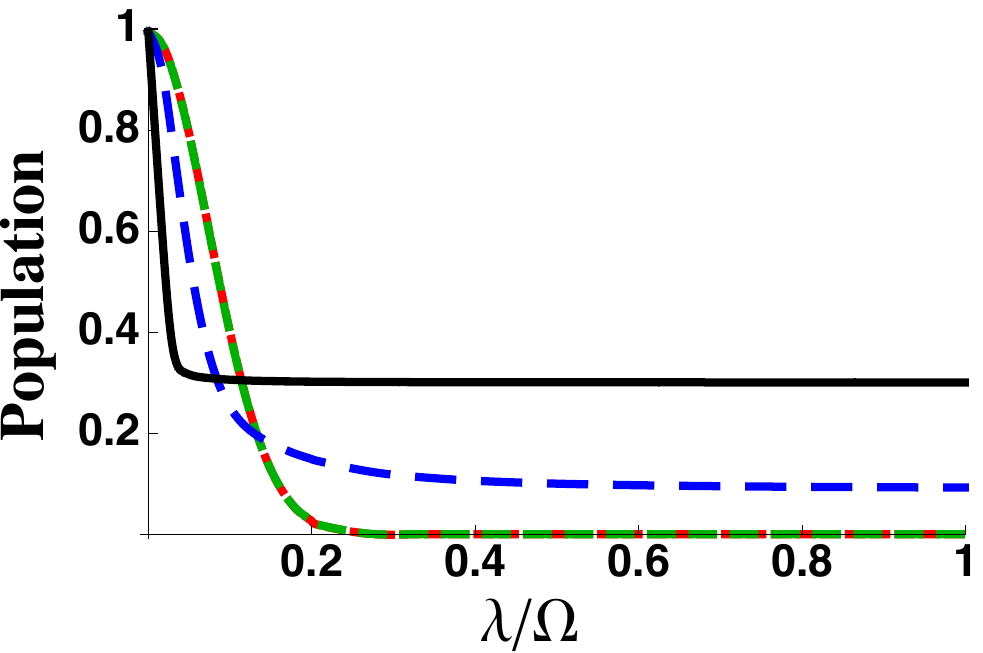}}
\caption{(Color online) Final population of state $\psiup_z$ when the system starts in $\psidown_z$ as a function of $\lambda/\Omega$, for different temperatures: $\kboltz T/\Omega = 0.001$ (red dotted line), $\kboltz T/\Omega = 1$ (green dashed line, almost superimposed to the red dotted one), $\kboltz T/\Omega = 3$ (blue long-dashed line) and $\kboltz T/\Omega = 10$ (black solid line). The relevant parameters are: $\kappa / \Omega^2 = 0.25$, $\kappa t_0 / \Omega = 25$, $\varphi=0$ and $\phi=\pi/2$ (which imply $\hat{X}=\hat{J}_x$).}\label{fig:temp-x}
\end{figure}

{\it Thermal effects ---} In Figures~\ref{fig:temp-z} and \ref{fig:temp-x} we show the effects of temperature. In Fig.~\ref{fig:temp-z} the efficiency of the population transfer as a function of $\lambda$ for different temperature is considered, when $\hat{X}=\hat{J}_z$. Higher values of temperature imply smaller values of efficiency.
Analogously, in Fig.~\ref{fig:temp-x} the efficiency is considered when $\hat{X}=\hat{J}_x$. 
It is interesting to observe that, in both cases, for very high temperature the value of the final population approaches a value close to $1/3$, which is well understood as a consequence of the occurring thermalization.

{\it Different Initial Condition --- } It is interesting to consider a different initial condition, say $\psiup_z$, which means that the adiabatic eigenstate carrying the population is $\psidown_\theta$, the instantaneous ground state of the system at every instant of time. On this basis, we can predict that, at zero temperature, the noise, and specifically the dissipation, can help the population transfer by inducing incoherent transitions from the higher levels to the lowest one.

\begin{figure}[h!]
\includegraphics[width=0.45\textwidth, angle=0]{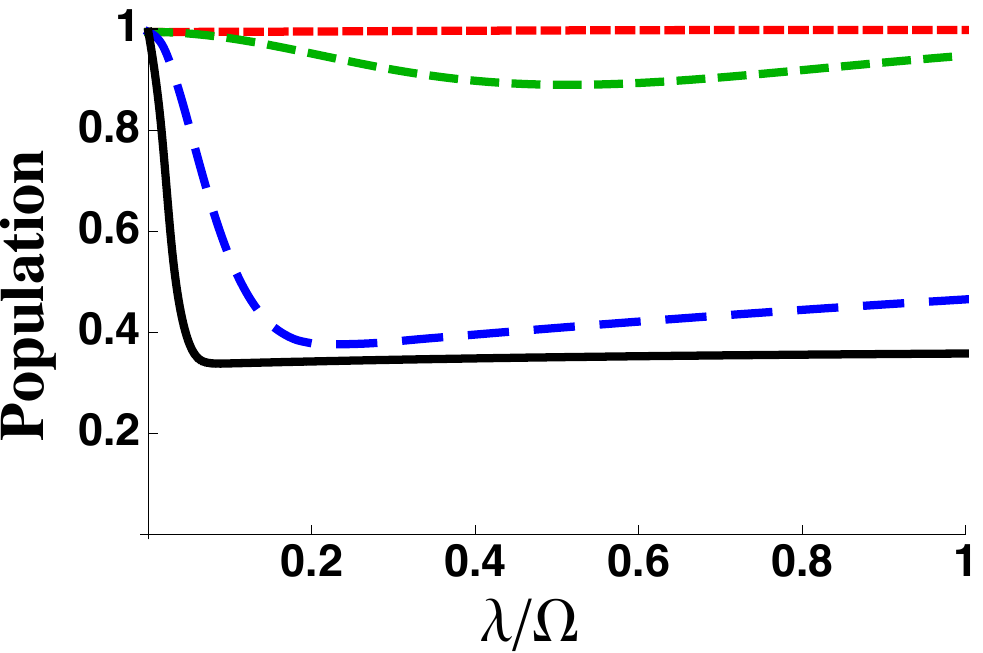}
\caption{(Color online) Final population of state $\psidown_z$ when the system starts in $\psiup_z$ as a function of $\lambda/\Omega$, for different temperatures: $\kboltz T/\Omega = 0.001$ (red dotted line), $\kboltz T/\Omega = 1$ (green dashed line), $\kboltz T/\Omega = 25$ (blue long-dashed line) and $\kboltz T/\Omega = 250$ (black solid line). The relevant parameters are: $\kappa / \Omega^2 = 0.25$, $\kappa t_0 / \Omega = 25$, $\varphi=0$ and $\phi=0$ (which imply $\hat{X}=\hat{J}_z$).} \label{fig:DiffInit-z}
\end{figure}

\begin{figure}[h!]
\includegraphics[width=0.45\textwidth, angle=0]{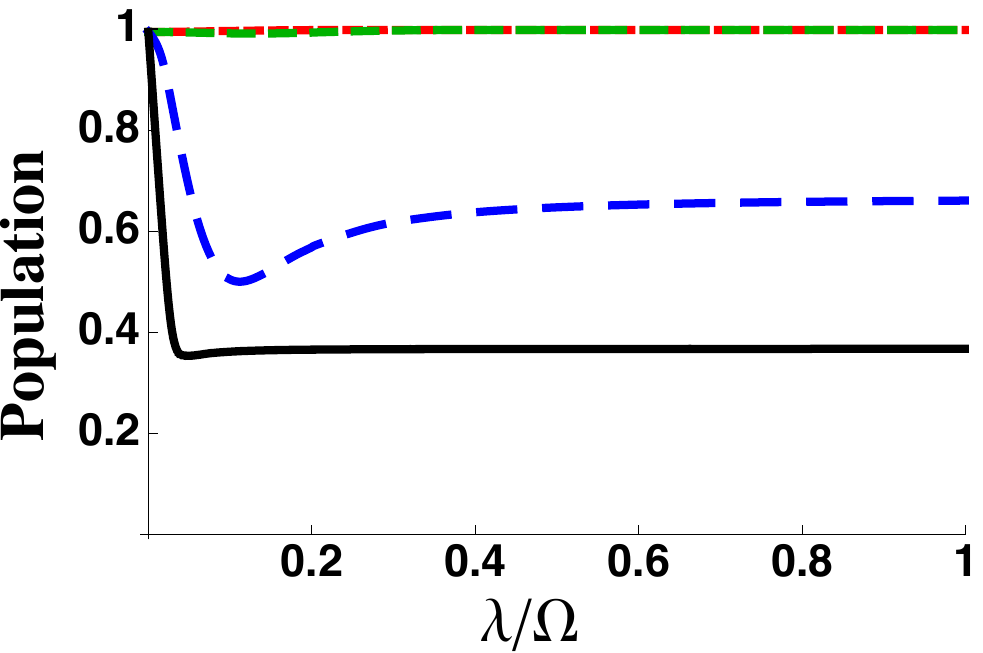}
\caption{(Color online) Final population of state $\psidown_z$ when the system starts in $\psiup_z$ as a function of $\lambda/\Omega$, for different temperatures: $\kboltz T/\Omega = 0.001$ (red dotted line), $\kboltz T/\Omega = 1$ (green dashed line, almost superimposed to the red dotted one), $\kboltz T/\Omega = 25$ (blue long-dashed line) and $\kboltz T/\Omega = 250$ (black solid line). The relevant parameters are: $\kappa / \Omega^2 = 0.25$, $\kappa t_0 / \Omega = 25$, $\varphi=0$ and $\phi=\pi/2$ (which imply $\hat{X}=\hat{J}_x$).} \label{fig:DiffInit-x}
\end{figure}

From Figures~\ref{fig:DiffInit-z} and \ref{fig:DiffInit-x} it is well visible that at essentially zero temperature (dotted red lines) the efficiency is very high. In the mainly dephasing case corresponding to $\hat{X}=\hat{J}_z$ (Fig.~\ref{fig:DiffInit-z}) the efficiency is also very high, though it admits some small deviations from unity. In the mainly dissipative case corresponding to $\hat{X}=\hat{J}_x$ (Fig.~\ref{fig:DiffInit-x}) the efficiency is essentially one for every values of $\lambda$.
In spite of having a different initial condition ($\Ket{1}_z$ instead of $\Ket{-1}_z$), also in these cases (both in mainly dissipative and mainly dephasing situations), for high temperature the asymptotic population approaches a value close to $1/3$ as an effect of the  thermalization process.

\section{Conclusions}\label{sec:conclusions}

In this paper we have analyzed the three-state Majorana model in the presence of interaction between the system and its environment. We have considered different cases, regarding the structure of the interaction Hamiltonian, identifying some special configurations which are the so called mainly dephasing case, for $\hat{X}=\hat{J}_z$, the mainly dissipative case, with $\hat{X}=\hat{J}_x$, and the purely dissipative case, corresponding to $\hat{X}=\hat{J}_y$. It is worth observing that, independently from their name, all such regimes are characterized by the presence of dissipation, at some level, and every dissipative dynamics is characterized by a loss of coherence.
Our analysis has been developed considering the generic mixed situation where the system-environment interaction involves a linear combination of the three fundamental angular momentum operators.

In the mainly dephasing model ($\hat{X}= \hat{J}_z$) the system exhibits a higher robustness against quantum noise than in the mainly dissipative case ($\hat{X}= \hat{J}_x$), as it is well visible from Fig.~\ref{fig:diss-1}.
On the other hand, the mainly dissipative model and the purely dissipative one ($\hat{X}\approx \hat{J}_y$) produce essentially the same effects on the efficiency, which is clearly shown by Fig.~\ref{fig:diss-2}, where the final population of the target state appears as perfectly independent from the angle $\varphi$. These assertions refer to a zero-temperature bath, but when the reservoir turns out to be at non-zero temperature, predictions change. In particular, at a very high temperature, in the presence of sufficiently high coupling constant $\lambda$, the environment is responsible for a thermalization process that brings the populations of the three adiabatic states of the system to almost the same value. Though one has a population of the target state higher than in the corresponding zero-temperature situation, one cannot see such an increase as a success of the protocol of population transfer, also considering the very low value reached ($\approx 1/3$).

The detrimental role of the environment can be turned into a positive factor when a different initial condition of the system is taken into account and the bath is at zero temperature. This is due to the fact that changing the initial condition implies changing the adiabatic state which carries population. If such state is the one with lowest energy then the zero-temperature quantum noise induces transitions toward the population carrier, therefore correcting the errors coming from diabatic transitions.

It is worth observing the crucial role of the chirping rate $\kappa$, as it emerges from Fig.~\ref{fig:kappa-z}. In an ideal situation, i.e., in the absence of quantum noise, a very low value of such parameter is preferable, since it allows to guarantee the validity of the adiabatic approximation, making the population transfer very efficient. Nevertheless, a lower $\kappa$ and the need to satisfy the condition $\kappa t_0\gg \Omega$ (to have the population carrier coincide with two bare states at the beginning and at the end of the process) imply a long duration of the experiment and hence a long exposure to the detrimental effects of the environment. Reaching an optimal value is a crucial point for the success of an experiment.

The types of decoherence discussed in this paper are relevant in various experimental situations. Though not exactly the same, pretty similar Hamiltonian models have been shown to describe specific experimental situation, such as two-qubit systems in the spin-$1$ subspace\cite{ref:Zhang2017} and three-state Rubidium atoms~\cite{ref:Song2016}, and in both experimental setups quantum noise is better taken into account.
Moreover, a possible different scenario is that of trapped-ions, for which decoherence shows up primarily as motional heating of the ion's phonon state.  Another example is found in ultracold atoms trapped in magneto-optical or dipole traps where the decoherence arises from the collisions between the atoms. For room-temperature atoms in a cell or in a beam decoherence is caused by both collisions and Doppler broadening. Yet another example is the classical NMR problem of a spin-1 in a magnetic field for which decoherence stems from the random fluctuations of the neighboring spins. In doped solids the decoherence is caused by the large inhomogeneous broadening and the fluctuations of the nuclear spins. In quantum dots decoherence is caused by the fluctuating electrostatic and spin environments as well as by the acoustic vibrations of the crystal lattice. The decoherence processes in all these and other systems involving three-state transitions can be described in the framework of the formalism used in this paper and therefore the present results are applicable to a large variety of experiments.


\section*{Acknowledgements}

NVV acknowledges support from ERyQSenS (Bulgarian Science Fund Grant No. DO02/3).


\end{document}